\documentclass[twocolumn]{aastex631}

\submitjournal{APJ}

\begin{document}

\title{Double-peaked Narrow Emission-Line Galaxies in SDSS-IV MaNGA}

\author[0000-0001-9410-9485]{Jiajie Qiu}
\affiliation{Shanghai Astronomical Observatory, Chinese Academy of Sciences, 80 Nandan Rd., Shanghai, 200030, China}
\affiliation{School of Astronomy and Space Science, University of Chinese Academy of Sciences, 1 East Yanqi Lake Rd., Beijing 100049, P.R. China}
\affiliation{Key Lab for Astrophysics, Shanghai, 200034, China}

\author[0000-0002-3073-5871]{Shiyin Shen}
\affiliation{Shanghai Astronomical Observatory, Chinese Academy of Sciences, 80 Nandan Rd., Shanghai, 200030, China}
\affiliation{Key Lab for Astrophysics, Shanghai, 200034, China}

\author[0000-0002-9767-9237]{Shuai Feng}
\affiliation{College of Physics, Hebei Normal University, 20 South Erhuan Road, Shijiazhuang 050024, China}
\affiliation{Guoshoujing Institute of Astronomy, Hebei Normal University, 20 South Erhuan Road, Shijiazhuang 050024, China}
\affiliation{Hebei Key Laboratory of Photophysics Research and Application, Shijiazhuang 050024, China}

\author{Yanmei Chen}
\affiliation{School of Astronomy and Space Science, Nanjing University, Nanjing 210093, China}
\affiliation{Key Laboratory of Modern Astronomy and Astrophysics (Nanjing University), Ministry of Education, Nanjing 210093, China}
\affiliation{Collaborative Innovation Center of Modern Astronomy and Space Exploration, Nanjing 210093, China}

\author[0000-0002-8733-1587]{Ruixiang Chang}
\affiliation{Shanghai Astronomical Observatory, Chinese Academy of Sciences, 80 Nandan Rd., Shanghai, 200030, China}
\affiliation{Key Lab for Astrophysics, Shanghai, 200034, China}

\author{Qianwen Zhao}
\affiliation{Shanghai Astronomical Observatory, Chinese Academy of Sciences, 80 Nandan Rd., Shanghai, 200030, China}
\affiliation{School of Astronomy and Space Science, University of Chinese Academy of Sciences, 1 East Yanqi Lake Rd., Beijing 100049, P.R. China}
\affiliation{Key Lab for Astrophysics, Shanghai, 200034, China}
\author{Qi Zeng}

\affiliation{Shanghai Astronomical Observatory, Chinese Academy of Sciences, 80 Nandan Rd., Shanghai, 200030, China}
\affiliation{School of Astronomy and Space Science, University of Chinese Academy of Sciences, 1 East Yanqi Lake Rd., Beijing 100049, P.R. China}
\affiliation{Key Lab for Astrophysics, Shanghai, 200034, China}

\correspondingauthor{Shiyin Shen}
\email{ssy@shao.ac.cn}

\begin{abstract}

Narrow emission lines in a galaxy's spectrum that show double peaks indicate the presence of distinct gas components with different velocities, and its physical origin remains uncertain. This study uses galaxies from the final MaNGA data release to detect double-peaked narrow emission-line spaxels (DPSs) by examining the double Gaussian profiles of the H$ \alpha $-[N \uppercase\expandafter{\romannumeral2}] doublets across all MaNGA spaxels. A total of 5,420 DPSs associated with 304 double-peaked narrow emission-line galaxies (DPGs) are identified, each DPG containing a minimum of 5 DPSs and being free from overlap with other galaxies. We find that DPSs can be categorized into three groups according to their central distance $r/R_e$ and the velocity difference $\Delta v$ between their two components: the inner low-$\Delta v$, inner high-$\Delta v$ and outer DPSs. By incorporating the physical characteristics of the DPGs into their DPSs, we demonstrate for the first time the existence of statistical correlations between barred DPGs and inner low-$\Delta v$ DPSs, AGN-hosting DPGs and inner high-$\Delta v$ DPSs, as well as tidal DPGs and outer DPSs.


\end{abstract}

\section{Introduction\label{sec:Introduction}}

There is a wealth of physical information contained within the optical spectra of galaxies \citep{Springel2005, Benson2010, Rubinur2016}. From the continua and absorption line features, the star formation history of galaxies can be decoded through stellar population synthesis techniques \citep{Conroy2009, Conroy2013}, whereas the emission lines also tell us various physical properties of ionized gas, e.g., gas-phase metallicity \citep{Tremonti2004, Kewley2019}, AGN activity \citep{Baldwin1981, Fernandes2001}, dust extinction \citep{Wang2004}, and gas kinematics \citep{Heckler2022}, etc.

Normally, a galaxy's spectrum only exhibits one set of emission lines, which characterizes the average feature of its ionized gas being observed. However, observations have shown that some galaxies display more than one set of lines \citep{Heckman1981, Heckman1984}, suggesting that there are multiple components with different systematic velocities \citep{Xu2009, Ge2012}. Galaxies with two sets of narrow emission lines are referred to as double-peaked narrow emission-line galaxies (DPGs) \citep{Wang2018}, which have been studied for decades \citep[e.g.,][]{Eracleous2012, Comerford2018}. Thanks to modern spectral surveys, such as the Sloan Digital Sky Survey \citep[SDSS,][]{Abazajian2009, Blanton2017} and the Large Sky Area Multi-Object Fiber Spectroscopy Telescope spectral survey \citep[LAMOST,][]{Shi2014, Wang2019}, the sample size of the DPGs has been significantly expanded. Furthermore, the physical origins of DPGs have also been explored in many studies from different points of view. For instance, \citet{Maschmann2020} performed a statistical study on a sample of 5,663 DPGs and found that there is an excess of lenticulars and a lack of late-type galaxies compared to normal galaxies. Additionally, based on the features of the velocity fields of different components, \citet{Nevin2016} summarized several mechanisms of the origin of DPGs, including inflows/outflows, spiral/bar structures, dynamic disturbance, and dust obscuration \citep[e.g.,][]{Eracleous1994, Liu2009, MullerSanchez2015}. Beyond that, double-peaked features may also be caused by double AGNs that are still merging \citep{Smith2010, Pilyugin2011}.

However, the intricate nature of the double-peaked emission lines makes it difficult to identify their physical origins without detailed data on the velocity fields of their two separate components. The solution to this issue lies in the use of the integral field spectroscopy (IFS) technique. For example, \citet{Wang2018} first investigated a galaxy using IFS data from Mapping Nearby Galaxies at APO \citep[MaNGA,][]{Bundy2014}, and inferred that the two sets of emission lines stem from a rotationally dominated structure and perturbation at the central region of the galaxy \citep{Nevin2016}. \citet{Yazeedi2021} studied the IFS observations of a galaxy with MaNGA ID 1-166919, a low-luminosity radio AGN, and proved that its double-peaked narrow emission-line components are associated with the bi-conical outflows of its central AGN. \citet{Ciraulo2021} analyzed an interesting target labeled as J221024.49+114247.0 also with MaNGA IFS data, and found that this target is a merging system in the pre-coalescence stage, and the two sets of emission lines are composed of two counter-rotating discs along the line of sight. Recently, \citet{Fu2023} conducted a systematic census of central broad-line AGNs and double-peaked emission lines to find double AGN candidates in the final release of MaNGA data \citep{Abdurro'uf2022}, in which 188 galaxies have been identified with their emission lines of central spaxels deviating significantly from single Gaussian profiles. Among these galaxies, 49 of them are considered to have double-peaked emission lines. 

As we have introduced, double-peaked emission lines are not exclusively found in the central regions of galaxies. This study will conduct a detailed analysis of these double-peaked narrow emission lines throughout each spaxel of every galaxy in the final release of the MaNGA data. Following this, we will compile a set of double-peaked narrow emission-line spaxels (DPSs) and identify their host galaxies as the DPGs. Our aim is to utilize this newly constructed statistical sample of DPSs and DPGs to enhance our understanding of the physical origins of the double-peaked emission lines.

The content of this paper is as follows. We describe the data sets and show the process of selecting the DPSs and DPGs in Section \ref{sec:Data}. We present the statistical properties of DPSs and DPGs in Sections \ref{sec:Results}. Using and combining this statistical information, we discuss the possible physical origins of DPSs and DPGs in Section \ref{sec:Discussion}. We finally make a summary in Section \ref{sec:Conclusions}.

\section{Data\label{sec:Data}}

\subsection{MaNGA Data Products\label{sec:MaNGA Data Products}}

As a large integral field spectroscopic survey, MaNGA is one of the core programs in the Fourth Generation Sloan Digital Sky Survey \citep[SDSS-IV,][]{Gunn2006, Blanton2017} that started in 2014 and ended in early 2021. About 10,000 nearby galaxies have been observed by hexagonal fiber-bundle integral field units \citep{Drory2015}. For each sample galaxy, the Data Reduction Pipeline \citep[DRP,][]{Law2016} of MaNGA provides the sky subtracted and flux calibrated 3D spectra named Cube, which has a wavelength coverage of 3,600 - 10,300 \AA, and a spectral resolution R $ \sim $ 2,000. By stacking and interpolation of several dithered exposures, the spatial resolution unit of the Cube is $0^{\prime\prime}.5 \times 0^{\prime\prime}.5$ and each resolution unit is named a spaxel. As an official software package for analyzing the Cubes, the Data Analysis Pipeline \citep[DAP,][]{Westfall2019} provides 2D Maps of various physical properties for each galaxy, e.g. the flux distribution and velocity field of emission lines. Both the DRP and DAP products can be accessed by Marvin \footnote{https://sas.sdss.org/marvin/} \citep{Cherinka2019}.

\subsection{Identification of DPSs\label{sec:Identification of DPSs}}

\begin{figure*}
\centering
\includegraphics[width=1\textwidth]{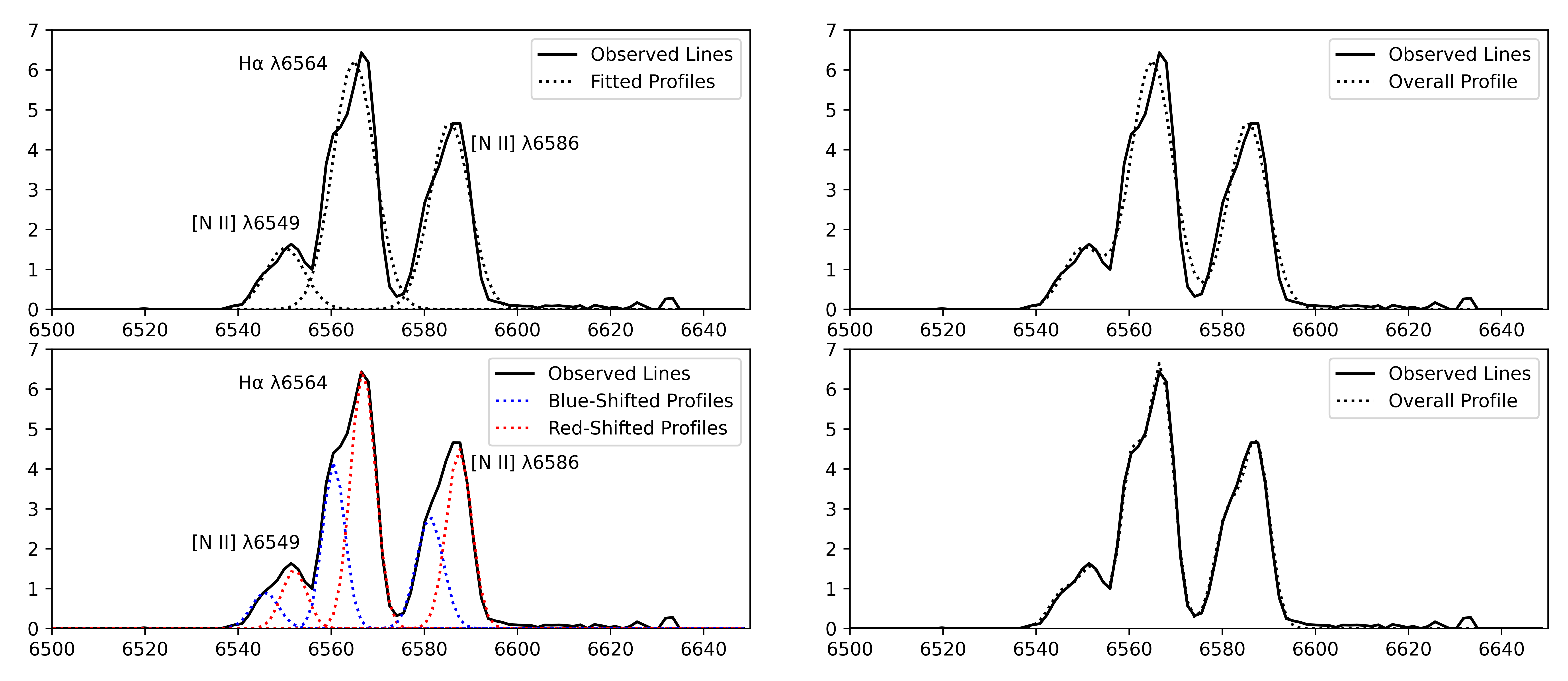}
\caption{\label{fig:SGFandDGF}
An example of the single and double Gaussian profiles fitting for the emission lines in a spaxel in the galaxy with MaNGA ID 1-593159. Top left: The original H$ \alpha $-[N \uppercase\expandafter{\romannumeral2}]$ \lambda\lambda $ 6549, 6586 emission lines are shown in the solid black profile in the wavelength ranging from 6,500 - 6,650 \AA. The black dotted line is the fitting result, containing 3 Gaussian profiles corresponding to [N \uppercase\expandafter{\romannumeral2}]$ \lambda$ 6549, H$ \alpha $ and [N \uppercase\expandafter{\romannumeral2}]$ \lambda$ 6586, respectively, from left to right. Top right: The overall profile is obtained from the result in top left panel. Bottom left: The solid black profile is the same as that in the top left panel. The dotted lines are also the fitting results, but it contains $2 \times 3$ Gaussian profiles. The blue and red lines represent two sets of emission lines blue- and red-shifted, respectively, each of which contains 3 Gaussian profiles corresponding to [N \uppercase\expandafter{\romannumeral2}]$ \lambda$ 6549, H$ \alpha $ and [N \uppercase\expandafter{\romannumeral2}]$ \lambda$ 6586 respectively from left to right.
Bottom right: The overall profile is obtained from the result in bottom left panel.
}
\end{figure*}

\begin{figure*}
\centering
\includegraphics[width=1\textwidth]{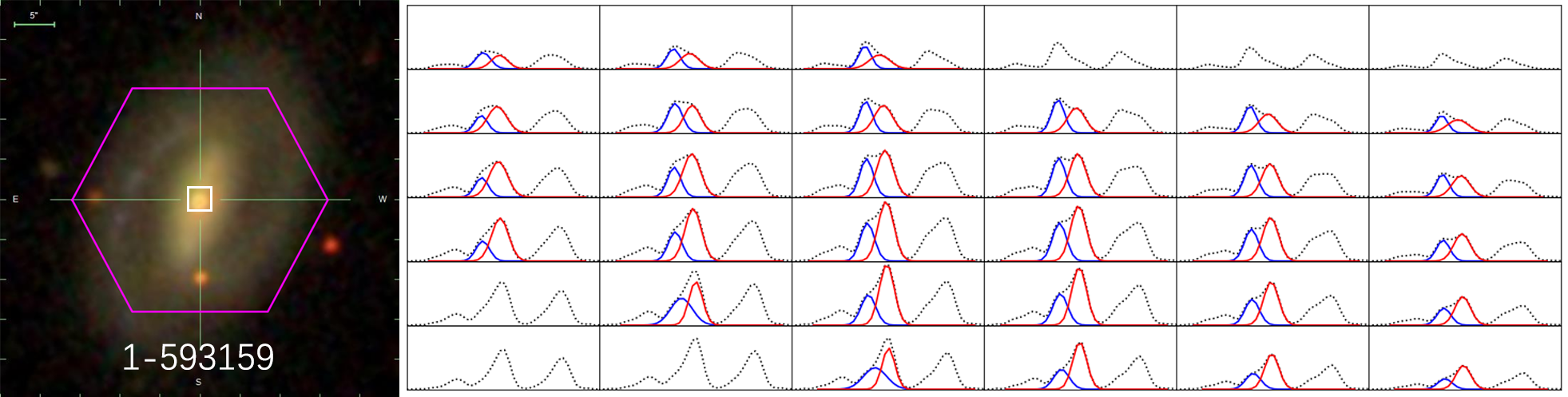}
\caption{\label{fig:DPS}
The SDSS image of the galaxy with MaNGA ID 1-593159 and the H$ \alpha $-[N \uppercase\expandafter{\romannumeral2}]$ \lambda\lambda $ 6549, 6586 line profiles of its central $6 \times 6$ spaxels ($3^{\prime\prime} \times 3^{\prime\prime}$, white box). All grids show the original emission lines with black dotted lines. Spaxels better fitted by the double Gaussian profiles are shown with the blue- and red-shifted H$ \alpha $ lines additionally.
}
\end{figure*}

There are 9,981 unique targets classified as galaxies in the final MaNGA data release. For each galaxy, we take the spectrum of each spaxel from its Cube and check whether it includes double-peaked narrow emission-line profiles. If so, we call this spaxel a DPS. Our processes are described in detail below.

First, we manually remeasure the emission-line parameters of each spaxel by subtracting the continua from the original spectra, provided by the 3D Cube for each galaxy. We use single and double Gaussian profiles to fit the combination H$ \alpha $-[N \uppercase\expandafter{\romannumeral2}]$ \lambda\lambda $ 6549, 6586 in the rest frame wavelength ranging from 6,500 - 6,650 \AA, with the nonlinear least-square method in the Python package scipy.optimize.curve\_fit \footnote{https://docs.scipy.org/doc/scipy/reference/generated /scipy.optimize.curve\_fit.html}. To ensure the reliability of this method, the average signal-to-noise ratio per wavelength unit in the wavelength range of each spaxel is required to be greater than 5 . For the single Gaussian profile fitting, there are 9 parameters, i.e., the amplitudes, velocities, and dispersions of the H$ \alpha $ and [N \uppercase\expandafter{\romannumeral2}]$ \lambda\lambda $ doublets. To reduce the free parameters, the velocities of the [N \uppercase\expandafter{\romannumeral2}] doublets are coupled with those of H$ \alpha $. The amplitude ratio and the dispersion ratio between [N \uppercase\expandafter{\romannumeral2}]$ \lambda $ 6586 and [N \uppercase\expandafter{\romannumeral2}]$ \lambda $ 6549 are fixed respectively at 3: 1 and 1: 1 \citep{DOJCINOVIC2022}. The remaining 5 free parameters are the amplitude, velocity, dispersion of H$ \alpha $ ($A_{\rm H\alpha}$, $v_{\rm H\alpha}$, ${\sigma}_{\rm H\alpha}$), and the amplitude and dispersion of [N \uppercase\expandafter{\romannumeral2}]$ \lambda $ 6586 ($A_{\rm [N \uppercase\expandafter{\romannumeral2}] \lambda 6586}$, ${\sigma}_{\rm [N \uppercase\expandafter{\romannumeral2}] \lambda 6586}$). For the double Gaussian profile fitting, we have 10 free parameters remained ($A_{\rm 1,H\alpha}$, $A_{\rm 2,H\alpha}$, $v_{\rm 1,H\alpha}$, $v_{\rm 2,H\alpha}$, ${\sigma}_{\rm 1,H\alpha}$, ${\sigma}_{\rm 2,H\alpha}$, $A_{\rm 1,[N \uppercase\expandafter{\romannumeral2}] \lambda 6586}$, $A_{\rm 2,[N \uppercase\expandafter{\romannumeral2}] \lambda 6586}$, ${\sigma}_{\rm 1,[N \uppercase\expandafter{\romannumeral2}] \lambda 6586}$ and ${\sigma}_{\rm 2,[N \uppercase\expandafter{\romannumeral2}] \lambda 6586}$) due to two sets of emission lines. We define the blue-shifted profiles as component 1 and the red-shifted ones as component 2 in each spaxel, that is, $v_{\rm 1,H\alpha}$ \textless $v_{\rm 2,H\alpha}$, for convenience in the following analysis. 

To effectively distinguish two velocity components, we require the velocity difference $\Delta v\equiv v_{\rm 2,H\alpha} - v_{\rm 1,H\alpha}$ with a lower limit of $\sim 200$ km/s (3 times the spectral resolution of MaNGA). Considering the difference between H$ \alpha \lambda$ 6564 and [N \uppercase\expandafter{\romannumeral2}]$ \lambda $ 6586 and in order to prevent confusion, we also set an upper limit of $\Delta v\sim 1000$ km/s. In this study, since we are interested in the narrow line components only, we require the velocity dispersion of both components (${\sigma}_{\rm 1,H\alpha}$ and ${\sigma}_{\rm 2,H\alpha}$) to be less than 600 km/s. Moreover, to ensure that the two identified components are physically comparable, we require the flux ratio between two components to be in the range $ 1/5 < f_{\rm 2,H\alpha} / f_{\rm 1,H\alpha} < 5$, calculated through $f_{\rm 2,H\alpha} / f_{\rm 1,H\alpha} \equiv (A_{\rm 2,H\alpha} \times {\sigma}_{\rm 2,H\alpha}) / (A_{\rm 1,H\alpha} \times {\sigma}_{\rm 1,H\alpha})$. 

After fitting the emission lines, we use the Bayesian Information Criterion \citep[BIC,][Equation \ref{BIC}]{Liddle2007} and the F test (Equation \ref{F-test}) together to identify DPSs. The BIC is defined as 

\begin{equation}
\Delta BIC = N_{\rm data} \times \ln{\frac{{\chi}_{\rm s}^2}{{\chi}_{\rm d}^2}} + (N_{\rm vars,s} - N_{\rm vars,d}) \times \ln{N_{\rm data}}\label{BIC}
\end{equation}
and the F test is 
\begin{equation}
F = \frac{{\chi}_{\rm d}^2 / N_{\rm vars,d}}{{\chi}_{\rm s}^2 / N_{\rm vars,s}}\label{F-test}\,.
\end{equation}

In both equations, $N_{data}$ is the number of sampling points in the wavelength range, while $N_{\rm vars,s} (= 5)$, ${\chi}_{\rm s}^2$ and $N_{\rm vars,d} (= 10)$, ${\chi}_{\rm d}^2$ are the numbers of free parameters and the minimum Chi-square values of the fitting of the single and double Gaussian profiles, respectively. We take the simultaneous fulfillment of the following two conditions as the criterion for identifying double-peaked emission lines: $ \Delta BIC > 0 $ \citep{Yazeedi2021} and $F > F_{0.05}(N_{\rm vars,s}, N_{\rm vars,d}) \sim 4.74 $. \footnote{ $F_{0.05}$ indicates a $5\%$ risk of concluding that a difference between single and double Gaussian profiles exists when there is no actual difference.} These spaxels with double-peaked narrow emission lines are defined as DPSs.

We display an example of H$ \alpha $-[N \uppercase\expandafter{\romannumeral2}] doublets in the wavelength range 6,500 - 6,650 \AA \, of a spaxel of the galaxy with MaNGA ID 1-593159 in Fig. \ref{fig:SGFandDGF}. Single and double Gaussian profiles are shown in the top and bottom panels, respectively. It is evident that, in the case of this spaxel, the asymmetry of the H$ \alpha $-[N \uppercase\expandafter{\romannumeral2}] doublets has been effectively addressed by the double Gaussian profiles, which provide a noticeably superior fit.

Due to the fact that the spaxels are the result of stacking and interpolation of several dithered exposures \footnote{The MaNGA integral field units are made from $2^{\prime\prime}.0$ core diameter fibers and with $0^{\prime\prime}.5$ gaps between adjacent fiber cores, and the minimum observing unit is a set of three dithered exposures for each MaNGA target.} and are not independent spatially resolved units, a real double-peaked feature would typically result in more than one DPS in the MaNGA data cube. Considering this, we only choose galaxies that have a minimum of 5 DPSs ($N_{\rm DS} \ge 5$) to be potential DPGs.

As an example, we show the image and spatial emission-line profiles for the DPG (MaNGA ID 1-593159) that contains DPSs in the right grids of Fig. \ref{fig:DPS}. The DPSs are located in the central $3^{\prime\prime}.0 \times 3^{\prime\prime}.0$ region (the white box on the left panel of Fig. \ref{fig:DPS}), where the doublets of H$ \alpha $ and [N \uppercase\expandafter{\romannumeral2}]$ \lambda\lambda $ are displayed for each DPS. In all of these grids, we plot the observed emission lines in each spaxel with black dashed lines. The fitted H$ \alpha $ lines of DPSs are shown with blue and red dotted lines, corresponding to the blue- and red-shifted profiles respectively.

\subsection{Final Sample of DPSs and DPGs\label{sec:Final Sample of DPSs}}

\begin{figure*}
\centering
\includegraphics[width=1\linewidth]{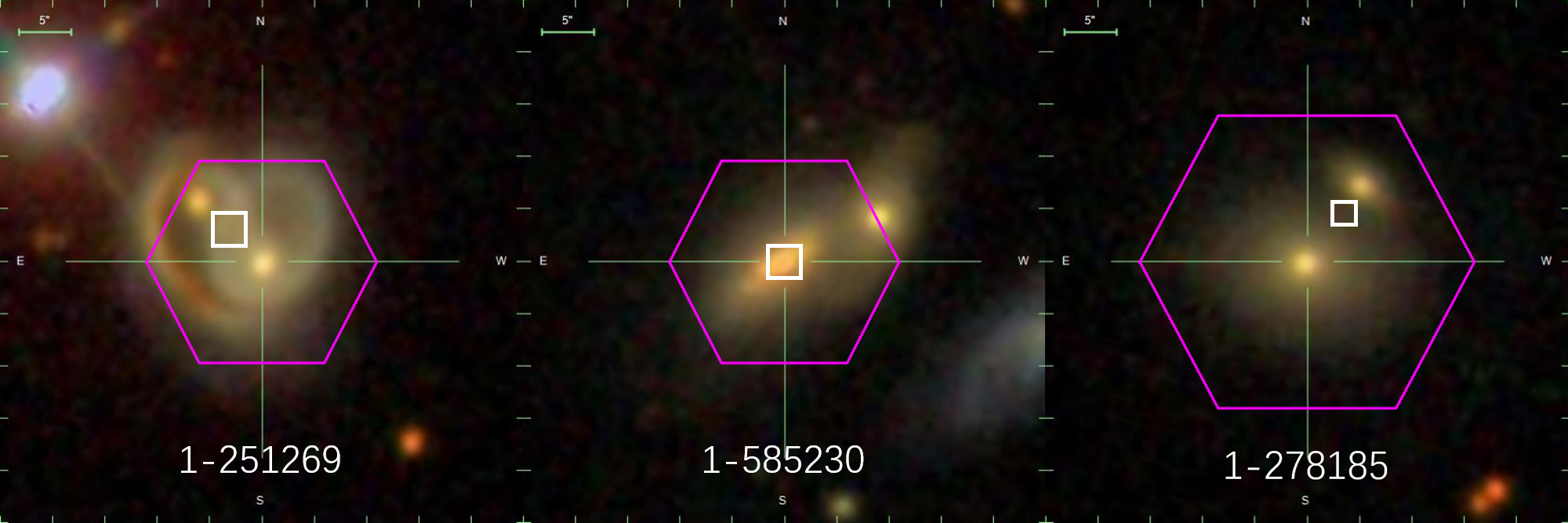}
\caption{\label{fig:MOS}
Three instances of paired galaxies exhibiting DPSs. In each galaxy, the location of the DPSs is roughly represented by a white rectangle. MaNGA ID 1-251269 (left) represents a merging system with prominent tidal features. MaNGA 1-585230 (middle) lacks distinct tidal features, yet the DPSs are positioned outside the overlapping area of the galaxy pair. MaNGA 1-278185 (right) displays a pair of galaxies without any tidal features, with DPSs located in the intersecting area of the two galaxies.}
\end{figure*}

Some of the DPSs may originate from the overlapping regions of galaxy pairs by projection effect \citep[e.g.][]{Ciraulo2021}, which is considered ordinary phenomena and is not of our interest in this study.   Through visual examination of the images of all DPG candidates, we identify 41 paired galaxies within the MaNGA field of view. Further analysis of the paired galaxies' morphology and the DPSs' positions reveals three distinct cases. In the first case, significant tidal features are shown in 14 paired galaxies (an example is illustrated in the left panel of Fig. \ref{fig:MOS}). Given that such merging activities could lead to the appearance of DPSs, these galaxies are retained in the DPG sample. In the second case, although no distinct tidal features are observed in the paired galaxies, the DPSs are not located in the overlapping areas of these galaxies. These 21 galaxies are maintained in the sample as well (an example is shown in the middle panel). The final case involves DPSs positioned directly within the overlapping zones of paired galaxies, which also lack notable tidal features. Recognized as a result of projection, this scenario leads to the exclusion of 6 galaxies from the final DPG sample (an example is shown in the right panel).

Finally, we obtain 5,420 DPSs and 304 DPGs. The results of these DPS are documented in Table \ref{tab:catalog1}. Beyond essential details like galaxy ID and DPS coordinates, we also detail the double-Gaussian fitting parameters of H$\alpha$ lines for each DPS.

\begin{table*}
\begin{center}
\begin{tabular}{l l l | r r r r r r | r r}
\hline\hline
MaNGA ID & x & y & $A_{\rm 1,H\alpha}$ & $v_{\rm 1,H\alpha}$ & ${\sigma}_{\rm 1,H\alpha}$ & $A_{\rm 2,H\alpha}$ & $v_{\rm 2,H\alpha}$ & ${\sigma}_{\rm 2,H\alpha}$ & r & $\Delta v$ \\
& & & $\rm 10^{-17}erg/$\AA$\rm /s/cm^2$ & (km/s) & (km/s) & $\rm 10^{-17}erg/$\AA$\rm /s/cm^2$ & (km/s) & (km/s) & ($R_e$) & (km/s) \\\hline
1-114955  &  37 &  40 &  2.87  &  -146.7 & 82.8 &  2.80 &  +81.4     &   124.5  & 0.35 & 228.0    \\
1-200058  &  29 &  36 &  2.23  &  -46.6 & 81.0 &  1.35 &  +200.2     &   101.7  & 0.62 & 246.8    \\
1-258380  &  26 &  26 &  2.19  &  -93.2 & 91.0 &  2.23 &  +129.8     &  108.5  & 0.11 & 223.0    \\
1-339094  &  17 &  18 &  1.65  &  -82.7 & 86.2 &  1.86 &  +163.2     &  115.1  & 0.22 & 245.9    \\
1-35268   &  22 &  20 &  1.11  &  -88.2 & 98.2 &  1.03 &  +167.3     &  113.2  & 0.16 & 255.5    \\
1-42250   &  21 &  21 &  1.82  &  -153.1 & 98.4 &  2.36 &  +152.7     &  118.9  & 0.07 & 305.7    \\
1-443132  &  26 &  27 &  5.78  &  -98.7 & 91.2 &  7.41 &  +141.7     &  93.2  & 0.15 & 240.4    \\
1-556749  &  22 &  23 &  2.46  &  -200.0 & 164.6 &  2.84 &  +184.2     &  152.4  & 0.14 & 383.9    \\
1-585711  &  24 &  40 &  0.43  &  +256.9 & 123.6 &  0.80 &  +564.9     &  113.1   & 0.57 & 308.0   \\
1-634140  &  20 &  34 &  0.41  &  -154.9 & 71.4 &  0.20 &  +203.4    &  133.5   & 0.91 & 358.3   \\
\hline
\end{tabular}
\caption{\label{tab:catalog1}Catalog of 5,420 DPSs in MaNGA galaxies. Column (1): the ID of the MaNGA galaxies. Column (2 and 3): the coordinates of the DPSs in MaNGA DAP. Column (4 - 9): Fitting parameters $A_{\rm 1,H\alpha}$, $v_{\rm 1,H\alpha}$, ${\sigma}_{\rm 1,H\alpha}$, $A_{\rm 2,H\alpha}$, $v_{\rm 2,H\alpha}$, ${\sigma}_{\rm 2,H\alpha}$. Column (10): Normalized centric distance to the effective radius of its host galaxy $r/R_e$. Column (11): Velocity difference $\Delta v$. The velocity dispersion for the two narrow components remains unadjusted for instrumental broadening.}
\end{center}   
\end{table*}

\section{Results\label{sec:Results}}

In this section, we initially conduct a statistical examination of the physical properties of DPSs, followed by an investigation into the attributes of their host DPGs. By associating these properties, we investigate the potential physical origins of various DPS types from a statistical perspective.

\subsection{Statistical Properties of DPSs\label{sec:Statistical Properties of DPSs}}

\begin{figure*}
\centering
\includegraphics[width=1\textwidth]{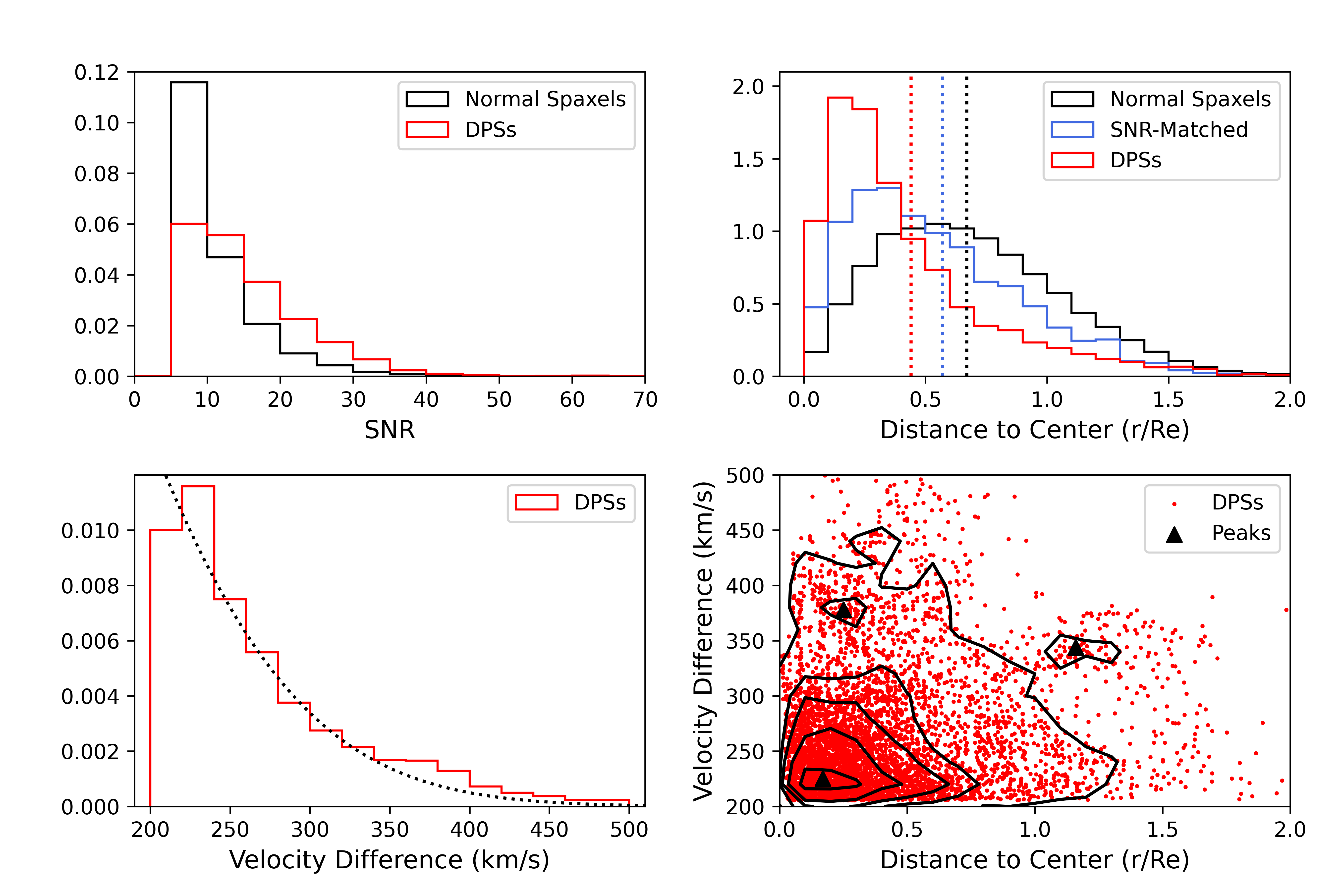}
\caption{\label{fig:snr_d}
Top left: The SNR ($> 5$) distributions of 5,420 DPSs (red) and all other normal spaxels (black). Top right: Distributions of $r/R_e$ of all DPSs (red), normal spaxels (black), and SNR-matched normal spaxels (blue). The average values of them are marked with dotted lines with the corresponding colors. Bottom left: Distributions of $\Delta v$ of DPSs, fitted by a Gaussian profile (dotted line). Bottom right: The DPSs in the $\Delta v$ - $r/R_e$ plane. Contours are plotted by their number density. The peaks are located at [0.17, 225], [0.25, 378] and [1.16, 344] in units [$R_e$, km/s] respectively.
}
\end{figure*}

For 5,420 DPSs, in addition to the fitting results obtained in Section \ref{sec:Identification of DPSs}, we also take their effective radii from the MaNGA DAP, which contains basic information on where the DPS features occur. For each spaxel, the normalized radius is defined as the normalized centric distance to the effective radius of its host galaxy, $r/R_e$.

We show the distributions of the SNR \footnote{The SNRs of spaxels is obtained by calculating the average signal-to-noise ratio of the emission lines between 6,500 - 6,650 \AA.} and $r/R_e$ of all 5,420 DPSs as red histograms in the top left and top right panels of Fig. \ref{fig:snr_d} respectively. For comparison, the SNR and $r/R_e$ distributions of all other normal spaxels in these DPGs are plotted as black histograms. Since there is an SNR cut to identify DPSs, we only compare spaxels with SNR $ > 5$. The average SNR of all DPSs is 15.3, higher than that of normal spaxels (11.0) in DPGs. Moreover, the DPSs are also biased toward inner regions of galaxies (0.44 $R_e$ VS 0.67 $R_e$). This finding suggests that either the DPSs are more easily detectable in spaxels with higher SNR, or that the spatial distribution of DPSs is biased towards the central regions, or both. To validate this, we create a control sample of spaxels by matching the SNR of normal spaxels to that of DPSs in the same galaxy on a one-to-one basis.

As can be seen in the top right panel of Fig. \ref{fig:snr_d}, the average $r/R_e$ of the SNR-matched normal spaxels is 0.57, still significantly larger than that of DPSs. This result indicates that the double-peaked features are indeed more likely to be connected to the activities of galaxies in the inner regions of galaxies. However, the occurrence of double-peaked features is not limited to nuclear regions. We have observed that approximately 10\% of our DPSs are located at distances greater than 1 $R_e$. This finding further supports the importance of using IFS instead of single-fiber spectroscopy for the identification of double-peaked features, as demonstrated in this study.

In addition to the SNR and the normalized centric distance, we have reliably measured the velocity difference $\Delta v$ of each DPS (see Section \ref{sec:Identification of DPSs}). We plot the histogram of $\Delta v$ in the bottom left panel of Fig. \ref{fig:snr_d}, which shows a wide distribution in the range of 200 to 500 km/s. In general, the number of spaxels decreases with increasing $\Delta v$, showing a Gaussian-like profile. We fit the $\Delta v$ distribution with a Gaussian profile (the mean value is fixed at 0) and get a result with the scatter of 135 km/s, which is plotted as a dotted line in the bottom left panel. Compared with the Gaussian profile, there is clearly an excess of a component at $\Delta v \sim$ 375 km/s.

To further explore the physical properties of DPSs with different values of $\Delta v$, we locate DPSs in the $\Delta v$ - $r/R_e$ plane in the bottom right panel of Fig. \ref{fig:snr_d}, where the contours of the number density of DPSs are also plotted for clarity. As can be seen, most DPSs are located at $\Delta v < $ 300 km/s and $r/R_e < $ 0.5, and peaked at $\Delta v \sim 220$ km/s and $r/R_e \sim 0.15$. In addition, there are two minor gathering areas: one is with high $\Delta v $ but still in the inner galactic regions ($\Delta v \sim $ 380 km/s, $r/R_e \sim 0.25$), and the other is also with high $\Delta v$ but in the outer regions ($\Delta v \sim $ 350 km/s, $r/R_e \sim 1.10$). For convenience, we refer to the DPSs located around these three regions in the $\Delta v$ - $r/R_e$ plane as inner low-$\Delta v$, inner high-$\Delta v$ and outer DPSs, respectively. 

In the following sections, we will use the knowledge obtained from the physical properties of host DPGs to examine the underlying physical mechanisms that lead to these three different categories of DPSs.

\subsection{Statistical Properties of DPGs\label{sec:Statistical Properties of DPGs}}

For each DPG, we obtain its physical properties mainly from the MaNGA DAP \citep{Westfall2019} and Value-Add Catalogs, such as the pipe3D Catalog \citep{Sanchez2016}. Other properties are obtained from additional sources, which will be referenced when utilized.

\subsubsection{Control Sample of DPGs\label{sec:Control Sample of DPGs}}

\begin{figure}
\centering
\includegraphics[width=.5\textwidth]{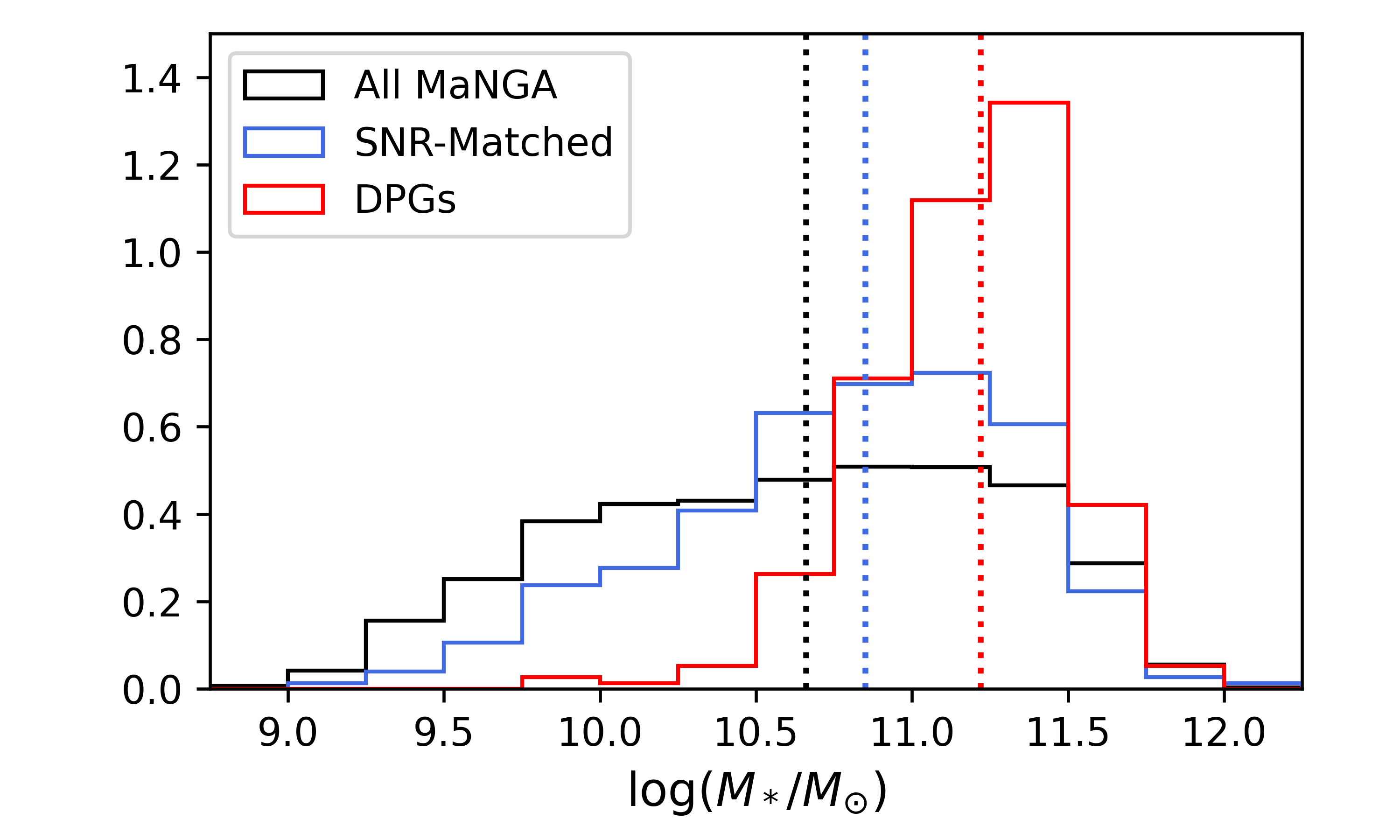}  
\caption{\label{fig:stellarmass}
The distributions of $M_*$ of all MaNGA galaxies (black histograms), DPGs (red), and the control sample of normal galaxies (blue). The mean values of the physical parameters are marked as vertical lines in the same color as the histograms.
}
\end{figure}

We first compare the distributions of the stellar mass ($M_*$, obtained from pipe3D) for DPGs and all MaNGA galaxies in Fig. \ref{fig:stellarmass}. To account for potential SNR bias caused by the selection of DPSs, we also create a control sample of normal galaxies with the same distribution of SNRs as the DPGs. We calculate the average SNR of all its spaxels with SNR $ > 5$ (obatin in Section \ref{sec:Statistical Properties of DPSs}) in each galaxy, as a representation of the SNR of each galaxy.

Compared to all MaNGA galaxies ($9.0 < \log M_* / M_\odot < 12.0$), the DPGs are distributed in higher mass range ($10.5 < \log M_* / M_\odot < 12.0$). Even after considering the selection bias from SNRs (blue histograms), DPGs are still significantly biased towards massive galaxies. This result is consistent with early findings from single-fiber spectroscopy, such as \citet{Maschmann2020}. The bias of DPGs toward massive galaxies might be related to the physical properties of galaxies (e.g. AGN phenomena). However, the bias in $M_*$ of DPGs could also be caused by the limited resolution of the MaNGA spectra and the ability of our DPS selection algorithm. For example, if the velocity of the peculiar component of a galaxy is similar to its circular velocity, our algorithm might fail to detect it for low-mass galaxies, which requires a minimum $\Delta v$ of 200 km/s. To rule out the effect of stellar mass and gain a better understanding of the physical properties of DPGs, we build a control sample of normal galaxies with the same $M_*$ and SNR distributions as DPGs.  By comparing with this control sample, we further explore the possible unique physical properties of DPGs in the subsequent detailed analysis.

\subsubsection{Physical Properties of DPGs\label{sec:Physical Properties of DPGs}}

\begin{figure*}
\centering
\includegraphics[width=1\textwidth]{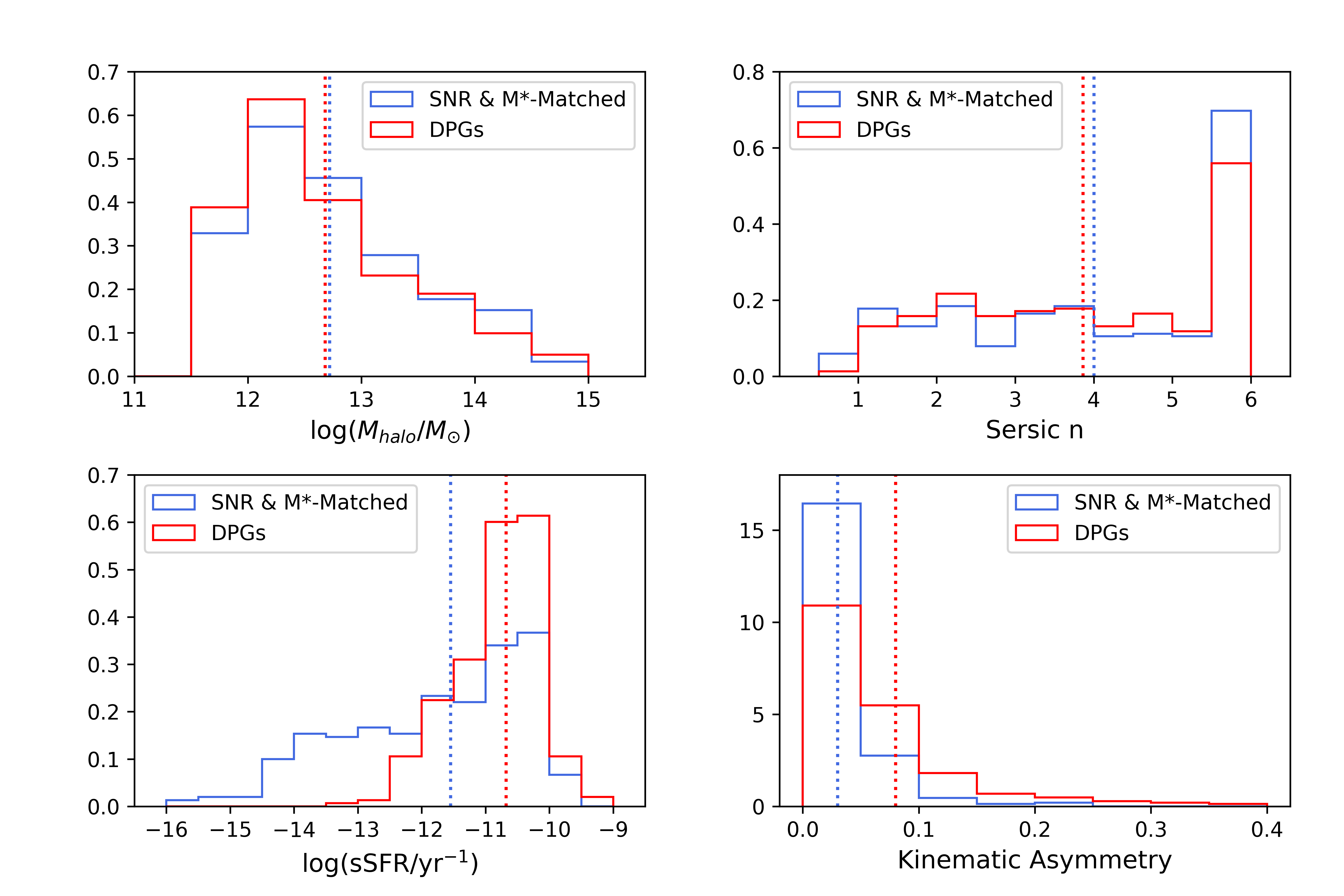}  
\caption{\label{fig:GlobalProperties}
The distributions of global properties of DPGs (red) and the control sample of normal galaxies (blue), including $M_{\rm halo}$ (top left), Sersic index $n$ (top right), $sSFR$ (bottom left) and Kinematic Asymmetry (bottom right).
}
\end{figure*}

We start from the global parameters of DPGs, including the host halo mass $M_{\rm halo}$, the Sersic index $n$, the specific star formation rate $sSFR$, and the Kinematic Asymmetry (KA) of the gas velocity fields. For each galaxy, the halo mass $M_{\rm halo}$ is derived using the luminosity $L_{19.5}$, as listed in the group catalog by \citet{Yang2007}. The Sersic $n$ values are listed in the MaNGA catalog, while the $sSFR$ of each galaxy is calculated by the $M_*$ and the global star formation rate (from pipe3D). The KA parameter is taken from \citet{Feng2022}, which is calculated as the ratio of the non-rotating higher-order Fourier components of the velocity fields of H$\alpha$ emission lines to the first-order component representing rotation inside the effective radius $R_e$.

In Fig. \ref{fig:GlobalProperties}, the distributions of the parameters of the DPGs and the control galaxies are examined across four panels. The top two panels show that there are no notable differences in the $M_{\rm halo}$ and $n$ distributions between the DPGs and the control galaxies, which implies that the presence of DPSs in galaxies is not directly related to either their host halo or global morphology. However, the DPGs shown in the lower two panels consistently demonstrate elevated levels of $sSFR$ and KA values relative to the control galaxies, even if the SNR impact is also corrected. These increased $sSFR$ and KA values seen in DPGs are consistent with the result of \citet{Feng2020}, which indicated that the galaxies with higher KA values show a notable increase in $sSFR$. However, as noted by \citet{Feng2022}, the high KA values in galaxies are not exclusively linked to galaxy interactions, but also to other physical features such as bar components. To further explore the physical origins of DPSs, we examine the features of the DPGs that relate to specific structures, including bars, AGNs, and tidal features.

\begin{figure}
\centering
\includegraphics[width=.5\textwidth]{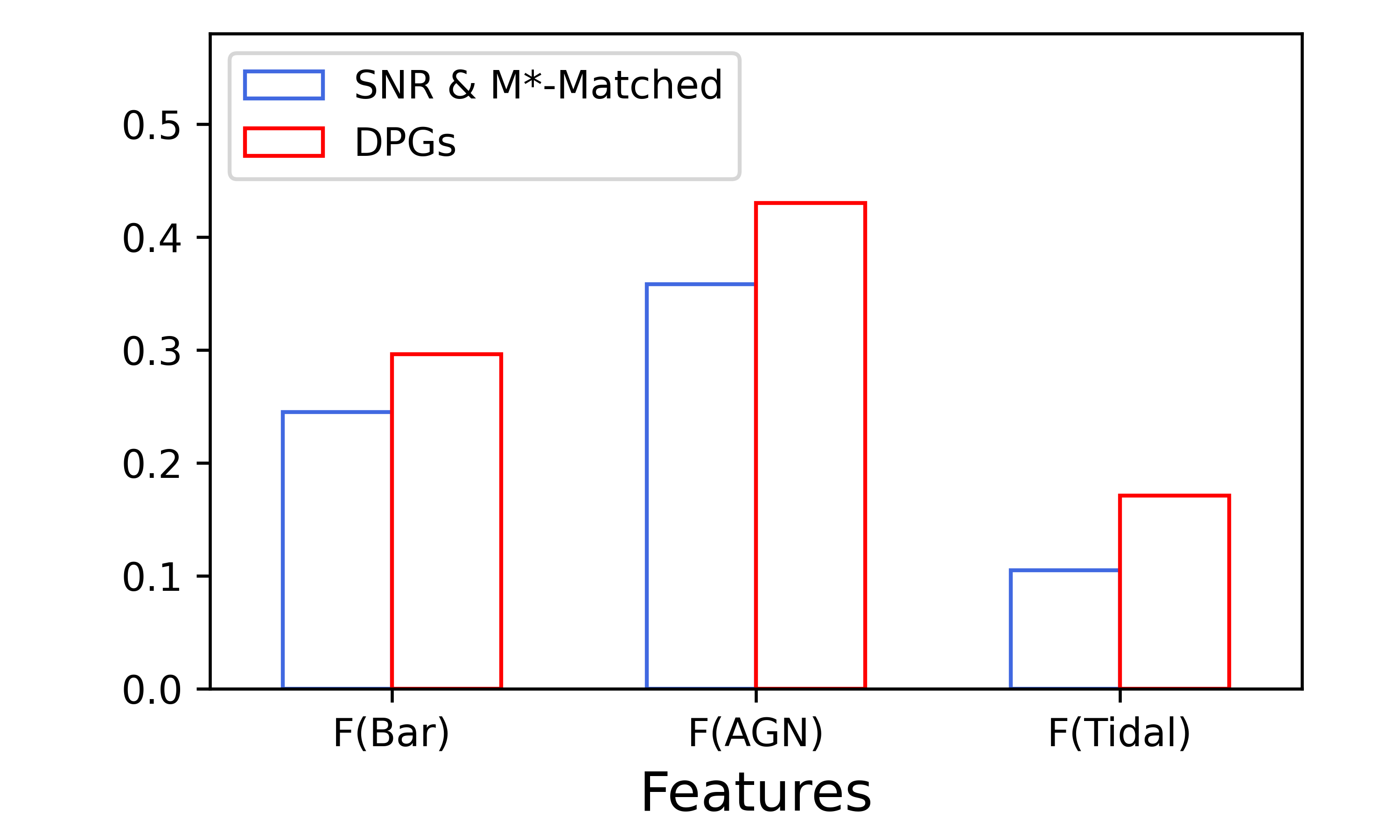}  
\caption{\label{fig:LocalFeatures}
The distributions of physical features of DPGs (red) and the control sample of normal galaxies (blue), including bars, AGNs, and tidal features, labeled as F(Bar), F(AGN), and F(Tidal), respectively.
}
\end{figure}

\begin{table*}
\begin{center}
\begin{tabular}{l | r r | r r r r}
\hline\hline
Types & All DPG & Control DPG & All DPS & Inner low-$\Delta v$ DPS & Inner high-$\Delta v$ DPS & Outer DPS \\
(Num) & (304) & (304) & (5,420) & (3,522) &  (1,040) & (858) \\\hline 
Bar & 0.296 & 0.245 & 0.307 &\textbf{0.363} & 0.154 & 0.260 \\
AGN & 0.430 & 0.358 & 0.480 & 0.422 & \textbf{0.675} & 0.484 \\
Tidal & 0.171 & 0.105 & 0.271 & 0.204 & 0.362 & \textbf{0.437} \\
\hline
\end{tabular}
\caption{\label{tab:catalog2}The number fractions of bars, AGNs, and tidal features for DPGs and DPSs. 
Column (1): the types of features. Column (2 and 3): the fractions of all DPGs and their control sample with these features. Column (4): the fractions of all DPSs with these features. Column (5 - 7): the fractions of sub-type DPSs with these features.}
\end{center}
\end{table*}

The bars and tidal features come primarily from the Galaxy Zoo 2 catalog \citep{Willett2013}, which contains 304,122 galaxies and covers more than 96\% of MaNGA galaxies. We consider a galaxy to have a bar or tidal feature when more than 60\% of the votes 'YES' are given. For the small number of galaxies that do not have a match in the Galaxy Zoo 2 catalog, we assess the presence of bars or tidal features through our own visual inspection. 

For AGN identification, we perform in both optical and radio bands. In optical, for each DPG, we use the parameters of the emission lines in the normal spaxels (i.e. all DPSs are excluded) that are the nearest to the galactic optical center (within $3^{\prime\prime}$) in DAP, and calculate the [N \uppercase\expandafter{\romannumeral2}]$/$H$\alpha$, [O \uppercase\expandafter{\romannumeral3}]$/$H$\beta$ and [S \uppercase\expandafter{\romannumeral2}]$/$H$\alpha$ flux ratios. We use [N \uppercase\expandafter{\romannumeral2}]- and [S \uppercase\expandafter{\romannumeral2}] - BPT diagrams \citep{Baldwin1981, Kewley2002} to identify the ionization mechanism of these spaxels, and divide their host DPGs into AGNs according to the criteria in \citet{Kewley2006}:
\begin{equation}
\rm 0.61 / ( log([N\ \uppercase\expandafter{\romannumeral2}] / H \alpha) - 0.47) + 1.19 < log( [O\ \uppercase\expandafter{\romannumeral3}] / H \beta)
\end{equation}
and
\begin{equation}
\rm 0.72/(log([S\ \uppercase\expandafter{\romannumeral2}]/H\alpha) - 0.32) + 1.30 < log([O\ \uppercase\expandafter{\romannumeral3}] / H\beta)
\end{equation}
Among the 304 DPGs examined, we identify 97 optical AGNs. In the radio band, the DPGs are compared with the AGN catalog of \citet{Best2012}, which includes 18,286 radio AGNs and encompasses the complete set of MaNGA galaxies. Among the 304 DPGs, we get 47 matched radio AGNs. Combining optical and radio AGNs, we get a total sample of 127 AGNs. Due to the relatively small number of radio AGNs,  no distinction is made between two types of AGNs in this study.

The number fractions of DPGs that exhibit bar/AGN/tidal features are 29.6\%/43.0\%/17.1\%, respectively, all of which are notably greater than those of the control galaxies (24.5\%/35.8\%/10.5\%), respectively These proportions are shown in Fig. \ref{fig:LocalFeatures} and detailed in Columns 2 and 3 of Tab. \ref{tab:catalog2}.

\subsection{Categories of DPSs: Connection to DPG features\label{sec:Categories of DPSs: Connection to DPG features}}

In Section \ref{sec:Statistical Properties of DPSs}, we show that there are three groups of DPSs on the $\Delta v$ - $r/R_e$ plane. Later in Section \ref{sec:Statistical Properties of DPGs}, we find that the host galaxies of DPSs are biased toward massive star-forming galaxies with highly asymmetric gas velocity fields and also toward galaxies with bar, AGN and tidal features. In this section, we further check how these three groups of DPSs may be connected with distinct physical features of their host DPGs.

\begin{figure*}
\centering
\includegraphics[width=1\textwidth]{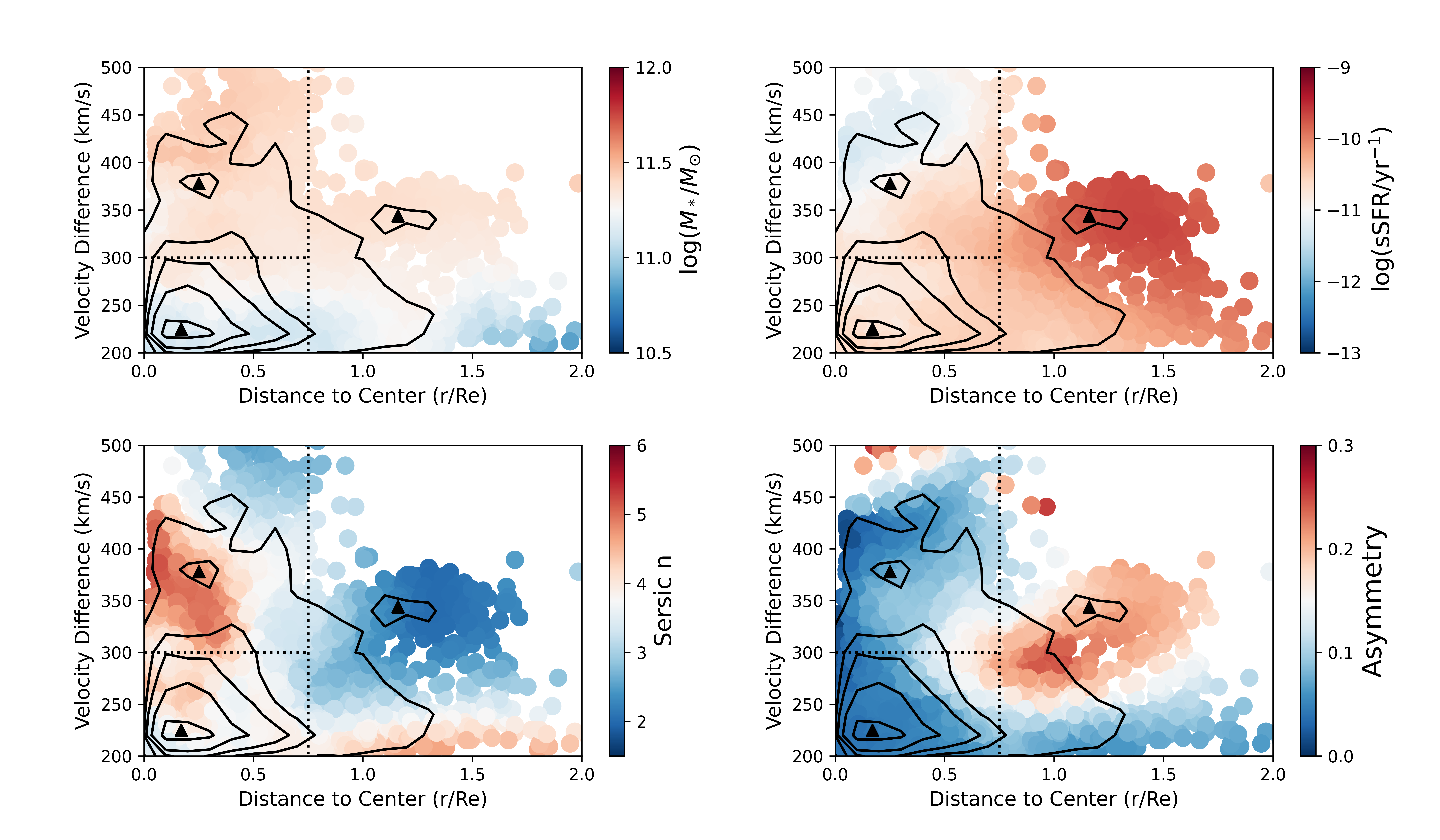}
\caption{\label{fig:d_dv1}
The color-coded distributions of the DPG parameters of the DPSs on the $\Delta v$ - $r/R_e$ plane, which have been smoothed by the LOESS algorithm. The DPG parameters include $M_*$ (top left), $sSFR$ (top right), Sersic index $n$ (bottom left), and Kinematic Asymmetry (bottom right). The number density of DPSs in the $\Delta v$ - $r/R_e$ plane is shown as the contours in each panel (the same as the bottom right panel of Fig. \ref{fig:snr_d}). Three types (the inner low-$\Delta v$, inner high-$\Delta v$ and outer DPSs) are divided by the dotted lines according to $r/R_e = 0.75$ and $\Delta v/(km/s) = 300$.
}
\end{figure*}

To facilitate a quantitative analysis, we categorize the DPSs into three groups on the $\Delta v$ - $r/R_e$ plane using the boundary lines $r/R_e =$ 0.75 and $\Delta v = $ 300 km/s, as shown by the dashed lines in Fig. \ref{fig:d_dv1}. This results in a total of 3,522 inner low-$\Delta v$ DPSs ($r/R_e < $ 0.75, $\Delta v <$ 300 km/s), 1,040 inner high-$\Delta v$ DPSs ($r/R_e < $ 0.75, $\Delta v > $ 300 km/s), and 858 outer DPSs ($r/R_e > $ 0.75).

We incorporate the position of each DPS in the $\Delta v$ - $r/R_e$ plane (Section \ref{sec:Statistical Properties of DPSs}) along with the physical parameters of its host DPGs (Section \ref{sec:Physical Properties of DPGs}). This process is carried out using the LOESS algorithm \citep{Cappellari2013}, which is a form of locally weighted regression. The results are then visualized using color codes on this plane. The corresponding plots are shown in the four panels of Fig. \ref{fig:d_dv1} for $M_*$, $sSFR$, Sersic index $n$, and KA, respectively.

The physical parameters of the host galaxies of three categories of DPSs exhibit significant differences. The inner low-$\Delta v$ DPSs are more commonly found in less massive, late-type galaxies (with low $M_*$, high $sSFR$, low $n$, and low KA values), whereas the inner high-$\Delta v$ DPSs are predominantly located in massive early-type galaxies (with high $M_*$, low $sSFR$, high $n$, and low KA values). The host galaxies of outer DPSs display the most distinctive features: high $M_*$, high $sSFR$, low $n$, and high KA values. 

These results clearly show that the different types of DPS that we categorized on the $\Delta v$ - $r/R_e$ plane are indeed related to different types of galaxies. Are they also associated with the three different physical processes that we have accounted, i.e. bars, AGNs and tidal features?

\begin{figure*}
\centering
\includegraphics[width=1\textwidth]{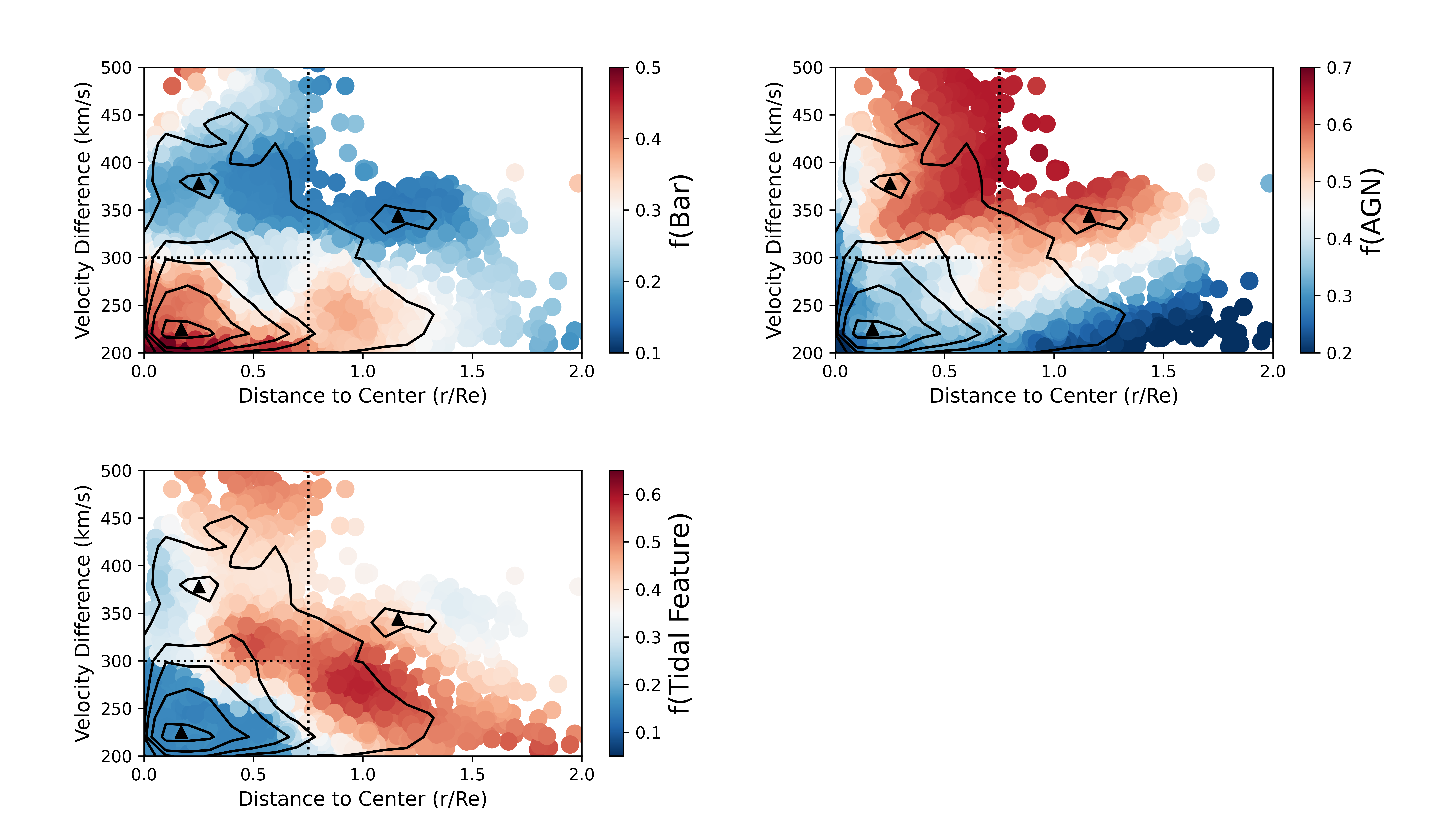}
\caption{\label{fig:d_dv2}
Similar plots as Fig. \ref{fig:d_dv1} but for physical features, including bars (top left), AGNs (top right) and tidal features (bottom left).
}
\end{figure*}

To verify this, we assign these three physical features of DPGs (yes or no) to each of their DPSs. Then, for all DPSs and each subset of DPSs, we calculate the fraction of them with each physical feature, respectively. We list the numbers of these fractions in the corresponding columns of Tab. \ref{tab:catalog2}.

For all DPSs (in Column 4 of Tab. \ref{tab:catalog2}),  30.7\% , 48.0\% and 27.1\% of them are in galaxies with bar/AGN/tidal features, respectively. These fractions are systematically higher than the global fractions of control samples with these features (24.5\% , 35.8\% and 10.5\%, in Column 3). This result reveals the fact that the number fraction of DPSs in barred/AGN-hosting/tidal DPGs is systematically higher than that of other type DPGs, which confirms the idea that the DPS phenomenon is intrinsically correlated with the bar/AGN/tidal features of galaxies.

For three subsets of DPSs, an interesting one-to-one relationship is observed with the bar/AGN/tidal features. As can be seen in Columns 5 - 7 of Tab. \ref{tab:catalog2}, the inner low-$\Delta v$, inner high-$\Delta v$ and outer DPSs have the highest fraction in barred, AGN host and tidal feature galaxies, respectively. As with Fig. \ref{fig:d_dv1}, the positions of DPSs in the $\Delta v$ - $r/R_e$ plane and their corresponding fractions with specific physical features are also presented in Fig. \ref{fig:d_dv2}.  There are evident one-to-one correlations between the three types of DPSs and three unique physical processes. The specifics of these correlations are enumerated and briefly discussed below.

\begin{itemize}
\item \textbf{Inner low-$\Delta v$ DPSs and barred DPGs}. Inner low-$\Delta v$ DPSs are the main type of our DPS sample (3,522 of 5,420). The number fraction of them in barred galaxies is 36.3\%, higher than the average value 30.7\%.  Bars are formed by the instability of galactic disks, which serve as a channel for gas inflow in galaxies \citep{MullerSanchez2015}. As illustrated by \citet{Maschmann2023}, the difference between the direction of the bar-induced gas inflow and the global rotation field of disk galaxies can lead to the emergence of the double-peaked features of emission lines. This scenario is consistent with the features of inner low-$\Delta v$ DPSs. 

\item \textbf{Inner high-$\Delta v$ DPSs and AGN-hosting DPGs}. In our 5,420 DPS sample, there are 1,040 inner high-$\Delta v$ DPSs. The number fraction of them in AGN-hosting galaxies is as high as 67.5\%, much higher than the average value 48.0\%. Early studies have shown that there is a strong correlation between AGN phenomena and bimodal features. As mentioned in \citet{KovaeviDojinovi2022}, AGN-hosting galaxies are found that the velocity differences of their double H$\alpha$ lines are from less than 200 km/s to over 700 km/s. \citet{MullerSanchez2015} examined a subset of 18 AGN-hosting galaxies with double-peaked narrow emission lines, in which 7 targets are considered to have AGN-driven bi-conical outflows by optical spectra, and 5 by both optical and radio methods. Considering that these outflows have velocities from several hundred to $\sim$ 1000 km/s lunched from the galactic center, this scenario is in good agreement with the appearance of the inner high $\Delta v$ DPSs in our study.

\item \textbf{Outer DPSs and tidal DPGs}. For 858 outer DPSs, the fraction of them in galaxies with tidal features is 43.7\%, while the average value is only 27.1\%. As shown by Fig. \ref{fig:d_dv1}, the host galaxies of these DPSs show clearly high $sSFR$ and kinematic asymmetry, consistent with the scenario of merging-induced DPSs \citep{Maschmann2020}. In our study, although we have removed paired galaxies with obvious overlap (Section \ref{sec:Final Sample of DPSs}) from our DPG sample, the case of DPSs induced by the projection effect of two paired galaxies cannot be entirely ruled out, such as MaNGA ID 1-114955 \citep{Ciraulo2021}.

\end{itemize}

\section{Discussion: Physical Origins of DPSs and DPGs\label{sec:Discussion}}

\begin{figure*}
\centering
\includegraphics[width=1\textwidth]{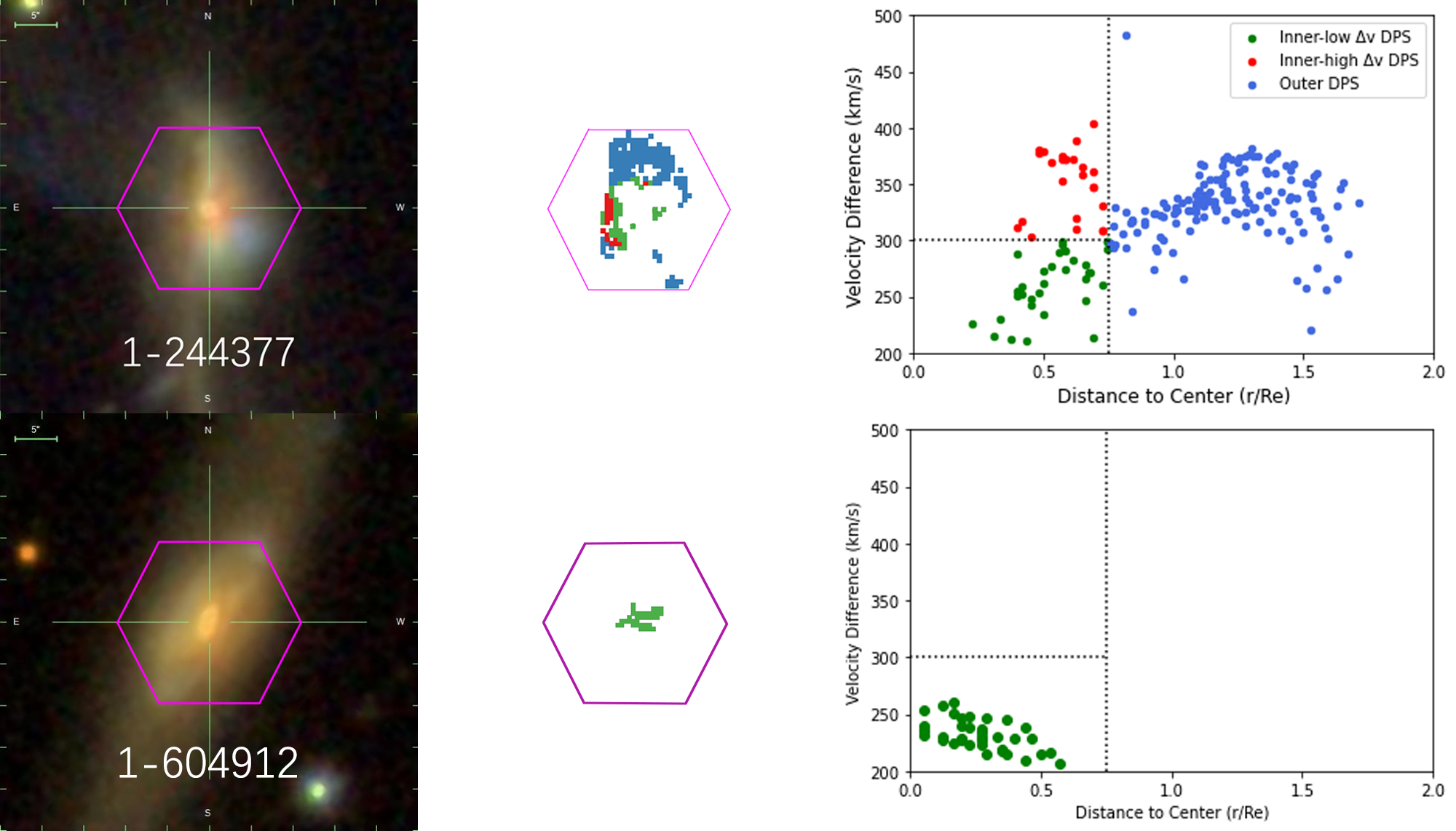}
\caption{\label{fig:example}
Two examples (MaNGA ID 1-244377 and 1-604912) of DPGs with different types of DPSs. The left panels shows their optical images with morphological features. The middle panels show the 2D distributions of different DPSs in the same fields of the left ones (Green: the inner low-$\Delta v$ DPSs; Red: the inner high-$\Delta v$ DPSs; Blue: the outer DPSs). The right panels give the 2D distributions of different DPSs in the $\Delta v$ - $r/R_e$ plane. MaNGA ID 1-244377 with a bar, an AGN and tidal feature has all three types of DPSs. MaNGA ID 1-604912 is an AGN-hosting galaxy, but only found inner low-$\Delta v$ DPSs at center.
}
\end{figure*}

In Section \ref{sec:Physical Properties of DPGs}, it has been demonstrated that there is evidence suggesting statistical connections between the inner low-$\Delta v$ DPSs and the barred DPGs, the inner high-$\Delta v$ DPSs and the AGN-hosting DPGs, and the outer DPSs and the tidal DPGs. However, there is no one-to-one connection between DPGs and their physical features, or DPGs and the categories of their DPSs. In other words, for a given DPG, it can contain more than one category of DPSs and have more than one physical feature (bar/AGN/tidal) probed in this study.  As an example, we show the galaxy MaNGA ID 1-244377 in the top panels of Fig. \ref{fig:example},  which includes all categories of DPSs and also is a galaxy confirmed to have a bar, an AGN and tidal feature (left panel). This result fully illustrates the complexity of the origin of multi-velocity components in the emission lines of galaxies. To better understand their physical origins, a more sophisticated study based on IFS data is clearly necessary and beneficial.

Moreover, we also emphasize that the connections between three categories of DPSs and bar/AGN/tidal features are statistical and can not be used as a criterion for identifying the physical origins of the DPS phenomena in any individual galaxies. The galaxy MaNGA 1-604912 is an example of an AGN-hosting galaxies. There is no obvious bar structure, but only has inner low-$\Delta v$ DPSs (bottom panels of Fig. \ref{fig:example}). In order to pinpoint the underlying physical causes of DPS phenomena in individual galaxies, it is necessary to incorporate more detailed physical information of the dual components. 

In fact, each DPS offers a wealth of additional information beyond $r/R_e$ and $\Delta v$ that can be derived from the emission lines themselves. For example, building the velocity and velocity dispersion fields of each component and exploring their connections with the global fields of normal spaxels could shed light on their physical natures \citep{Heckler2022}. Furthermore, if we can reliably identify the dual velocity components for multiple emission lines, it will be possible to derive other physical parameters (e.g., BPT diagram, Balmer decrement) for both velocity components. This approach would certainly contribute to the confirmation of the physical origin of DPSs \citep{Selwood2022}. Nevertheless, these detailed approaches tailored to individual galaxies have gone beyond the scope of this paper, a census and statistical study on all MaNGA spaxles, but will be the focus of our future study.

\section{Conclusions\label{sec:Conclusions}}

In this study, we use the complete dataset of 9,981 galaxies from the final release of the SDSS MaNGA survey to methodically detect double-peaked H$ \alpha $-[N \uppercase\expandafter{\romannumeral2}]$ \lambda\lambda $ 6549, 6586 lines within their spaxels. Spaxels exhibiting a velocity difference exceeding 200 km/s between the two peaks are identified as double-peaked spaxels (DPSs), and galaxies containing at least 5 DPSs are classified as double-peaked emission line galaxies (DPGs). After removing the projection effects from paired galaxies, we finally build up a sample of 304 DPGs, including a total of 5,420 DPSs. Through the analysis of DPS distributions on the plane defined by velocity difference ($\Delta v$) and central distance ($r/R_e$), it has been observed that DPSs cluster into three distinct groups. The primary group is found in the inner region ($r/R_e < 0.75$) with low-$\Delta v$ ($\Delta v < 300$ km/s), containing 3522 DPSs. The other two groups consist of the inner region with high-$\Delta v$ ($r/R_e < 0.75$, $\Delta v > 300$ km/s) and the outer region ($r/R_e > 0.75$), including 1040 and 858 DPSs, respectively.

Upon comparing a control sample of regular galaxies (lacking DPSs) that share the same stellar mass distribution, we find that the DPGs exhibit similar overall morphology (described by Sersic index $n$) and host halo mass distributions to the control galaxies. However, DPGs demonstrate increased specific star formation rates and higher kinematic asymmetry. Moreover, DPGs show a more frequent occurrence of galaxies with bars, AGNs, and tidal features. The key finding of this study is the statistical correlation between the three subgroups of DPSs and three unique galactic physical features(bar/AGN/tidal). Specifically, DPSs with lower inner $\Delta v$ are linked to bar structures, those with higher inner $\Delta v$ are related to AGNs, and the outer DPSs are connected to tidal structures. Despite this statistical correlation, for an individual DPG, we emphasize that the physical origin of its DPSs can be very complex.  A galaxy may have more than one type of DPS and more than one specific physical feature. There is not necessarily a one-to-one correspondence between them for individual galaxies. To obtain the physics of the origin of DPSs in a specific galaxy, it is necessary to introduce more observational features and construct more complex physical hydrodynamic models \citep[such as][]{Feng2024}, which we will do as follow-up works.

\section*{ACKNOWLEGEMENTS}

SS thanks research grants from the China Manned Space Project with NO. CMS-CSST-2021-A07, the National Key R\&D Program of China (No. 2022YFF0503402, 2019YFA0405501,), National Natural Science Foundation of China (No. 12073059 \& 12141302) and Shanghai Academic/Technology Research Leader (22XD1404200).

Funding for the Sloan Digital Sky Survey IV has been provided by the Alfred P. Sloan Foundation, the U.S. Department of Energy Office of Science, and the Participating Institutions. SDSS-IV acknowledges support and resources from the Center for High Performance Computing at the University of Utah. The SDSS website is www.sdss.org.

SDSS-IV is managed by the Astrophysical Research Consortium for the Participating Institutions of the SDSS Collaboration including the Brazilian Participation Group, the Carnegie Institution for Science, Carnegie Mellon University, Center for Astrophysics — Harvard \& Smithsonian, the Chilean Participation Group, the French Participation Group, Instituto de Astrofisica de Canarias, The Johns Hopkins University, Kavli Institute for the Physics and Mathematics of the Universe (IPMU) / University of Tokyo, the Korean Participation Group, Lawrence Berkeley National Laboratory, Leibniz Institut fur Astrophysik Potsdam (AIP), Max-Planck-Institut fur Astronomie (MPIA Heidelberg), Max-Planck-Institut fur Astrophysik (MPA Garching), Max-Planck-Institut fur Extraterrestrische Physik (MPE), National Astronomical Observatories of China, New Mexico State University, New York University, University of Notre Dame, Observatario Nacional / MCTI, The Ohio State University, Pennsylvania State University, Shanghai Astronomical Observatory, United Kingdom Participation Group, Universidad Nacional Autonoma de Mexico, University of Arizona, University of Colorado Boulder, University of Oxford, University of Portsmouth, University of Utah, University of Virginia, University of Washington, University of Wisconsin, Vanderbilt University, and Yale University.

\bibliography{reference}{}
\bibliographystyle{aasjournal}

\end{document}